\documentclass[english,twocolumn,showpacs,preprintnumbers,amsmath,amssymb]{revtex4}

\usepackage{graphicx,color,babel}

\def\be{\begin{equation}}
\def\ee{\end{equation}}
\def\bc{\begin{center}}
\def\ec{\end{center}}

\begin{document}

\title{Phase transition and phase coexistence in coupled rings with driven exclusion
process}

\author{Rakesh Chatterjee}
\author{Anjan Kumar Chandra}
\author{Abhik Basu}
\affiliation{Theoretical Condensed Matter Physics Division, Saha
Institute of Nuclear Physics, Calcutta 700064, India
\\
}

\date{\today}
\def \aa {\alpha_{_A}}
\def \ab {\alpha_{_B}}
\def \ba {\beta_{_A}}
\def \bb {\beta_{_B}}

\def\rd{\textcolor{black}}
\def\rr{\textcolor{black}}

\begin{abstract}
We study one-dimensional exclusion processes in two coupled closed
rings consisting of a common diffusive channel and two parallel
active (driven) channels. Our model displays bulk-driven phase
transition and phase coexistence in the form of a localised domain
wall (DW) in one of the active channels in a limit where the
diffusive and driven dynamics compete. By \rd{controlling a splitting
parameter which tunes the in-coming currents into the active
channels,} the system can be brought to a delocalisation transition,
when delocalised DWs are formed in both the active channels. We
characterise the DW fluctuations numerically.
\end{abstract}

\maketitle

\section{Introduction}

Totally Asymmetric Simple Exclusion Process (TASEP) \cite{tasep}
serves as a paradigmatic example of open non-equilibrium systems in
one dimension ($1d$). Its practical realizations include quasi-$1d$
motion of molecular motors along with microtubules in intra-cellular
transport \cite{how}, protein synthesis \cite{pro} or motion in
geometrical confinement, e.g., nuclear pore complex of cells
\cite{pore}. In contrast, Symmetric Exclusion Process (SEP)
\cite{sep} is a typical example of $1d$ diffusion. Well-known
examples of SEP include diffusion through artificial crystalline
zeolites \cite{zeo}. In both passive (SEP) and active (TASEP or
TASEP-like) systems, prohibition of mutual passage of particles or
{\em exclusion} gives rise to nontrivial collective effects, whose
details of course depend upon whether the dynamics in question is
TASEP or SEP. Active systems with open boundaries generally display
spatially nontrivial steady state density distributions.

In this paper, we propose a closed model that consists of two
overlapping rings with a common diffusive part (SEP) and two
parallel active (driven) channels (marked $T_A$ and $T_B$
hereafter). In order to ensure competition between driven and
diffusive dynamics, we consider a particular limit of the model. Our
principal result includes identification of a model parameter
$\theta$, having values between 0 and 1 (see below), as a switch, by
tuning which continuously keeping everything else unchanged (i) one
may de-pin pinned domain walls (DW) and (ii) \rd{ as $\theta$ crosses $1/2,$
a localised DW in one of the active channels disappears and appears in the other}.
Our model should serve as a paradigmatic example of
localisation-delocalisation transition in a $1d$ closed model with
coupled diffusive and driven dynamics. In addition to its direct
theoretical relevance, it is phenomenologically motivated by the
movement of molecular motors in closed compartments \cite{closemot},
the dynamics of colloidal particles in optical traps \cite{coll} and
the dynamics for multiple mRNAs competing for finite resources
(ribosomes), where the ribosomes in turn are bounded by a certain
trajectory and a diffusion rate outside the mRNA (during
recycling)~\cite{cook,protein_synthesis}. \rd{ In particular,
protein synthesis involves two stages: transcription of genetic
information from DNA to messenger RNA (mRNA) by RNA polymerase and
translation from mRNA to proteins through ribosome translocation. In
most bacteria such as E.coli, translation involves three main
players: the mRNA (genetic template), the ribosome (assembly
machinery), and aminoacyl transfer RNAs (aa-tRNAs), i.e., transfer
RNAs ``charged" with the corresponding amino acid. The process of
translation consists of ribosomes moving along the mRNA without
backtracking. This is modeled by TASEP. It is well-known that
ribosomes that move along mRNA strand are recycled in a cell. For
instance, in eukaryotic cells, after each round of protein
synthesis, the ribosomes are released from the mRNA and they join
the common pool of ribosomes in cytoplsm, where they execute
diffusion and may rejoin the mRNA to restart protein synthesis. In
our model, the SEP channel models the ``common pool" of diffusive
ribosomes in the cytoplasm of an eukaryotic cell, which in our model can come
back to the entry point of the TASEP lanes due to the feedback from
the SEP channel.} \rr{ However, although ribosome translocation along mRNA forms physical motivation of the
present work, the analogy between our model and the actual biological process of ribosome translocation along
mRNA strands is not strict due to various limitations of our model, as we discuss below.} In both SEP and TASEP, each lattice site has
maximum unit occupancy \rd{ and a particle can only move to the
nearest neighbour site (in both directions for SEP or in one
direction only for TASEP), only if that site is empty. Thus the
dynamics obeys the exclusion principle.} For SEP with open boundaries, the
density profile is always linear with the slope being determined by
the boundary conditions at the two ends \cite{sep}. In contrast
TASEP with open boundaries displays three distinct phases
\cite{tasep-phase} characterized by their average densities (low and
high) and a third phase marked by a maximal current (MC). Unlike
with open boundary conditions, individual SEP and TASEP dynamics
with closed boundaries (say, closed rings) exhibit only uniform
density profile in the steady state due to spatial translational
invariance. The rest of the paper is organised as follows: In
Sec.~\ref{model} we discuss our model in details. Then in
Sec.~\ref{steady} we set up our mean-field theory (MFT) and discuss
the steady state density profiles by using our MFT, complemented by
extensive Monte-Carlo simulation (MCS) studies. In Sec.~\ref{local}
we go beyond MFT, and discuss domain wall fluctuations and
delocalisation transition at $\theta=1/2$. Finally, in
Sec.~\ref{conclu} we summarize our results.

\section{The Model}
\label{model}
\begin{figure}[h]
\centering
\includegraphics[width=7.5cm,bb=14 14 415 215]{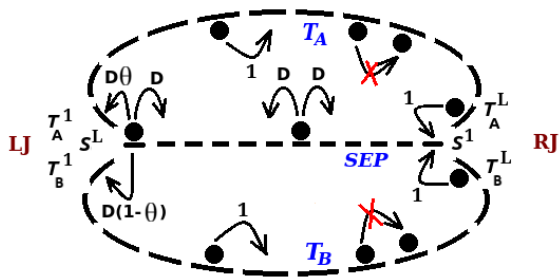}
\caption{\rr{(color online)} Schematic diagram of the model; LJ/RJ refer to the
left/right junctions. Site labels run from $m=1$ to $L$ from LJ to
RJ for $T_{A,B}$ and for SEP from RJ to LJ, superscripts denote 
site numbers for the corresponding channel $T_A$, $T_B$ and $S.$ }
\label{fig:model1}
\end{figure}
Our proposed $1d$ model is a closed system of two overlapping rings
consisting of three channels of equal number of sites designated by
$m=1,2,...L,$ as shown in Fig.~(\ref{fig:model1}). Dynamics of the
two channels $T_A$ and $T_B$ are governed by TASEP and the particles
in the third channel \rd{$S$} execute SEP.  \rd{Thus in $S$} particles can
hop to both direction with rate $D$, whereas in $T_A$ and $T_B$
particles can only hop to its right neighbour if empty with rate
unity setting the time scale. At the left junction of $T_A$ and $T_B$ particles can either
enter from SEP channel with rate $D\theta$ and $D(1-\theta)$
respectively if those sites are vacant or can hop to the other side
with rate $D$. \rd{ If both $T_A$ and $T_B$ try to inject a particle into
SEP, then one of the TASEP channels ($T_A$ or $T_B$) is selected
randomly for injecting a particle to the target site i.e, the first
site of SEP if it is empty.} If $N_p$ be the total number of particles then, the global particle density $n_p = N_p/3L.$ \rr{In Fig.~(\ref{fig:model1}), symbols $LJ$ and $RJ$ refer to the left and right junctions in the model.
In addition, for the purpose of clarity, the site number for a
particular channel (i.e., $T_A$, $T_B$ or $S$) are given as a superscript, e.g., at $LJ$, the first sites of $T_A$ and $T_B$ are denoted as $T_A^1$ and $T_B^1$, respectively, and the $L$-th site of $S$ is denoted as $S^L$. Similarly for the right junction $RJ$.} In this model, the three
bulk parameters $(\theta,n_p,D)$ control different phase
transitions. Thus we observe \textit{bulk induced} phase transitions
unlike the usual TASEP with open boundaries which display boundary
induced phase transition. The steady state current in each of SEP,
$T_A$ or $T_B$ is a function of $\theta,n_p,D$ and is spatially
constant. Notice that our model is a variant and extension of that
in Ref.~\cite{frey}. In particular, for $\theta=1$ or 0, $T_B$ or
$T_A$ is blocked and our model explicitly reduces to that of
Ref.~\cite{frey}. Evidently, for $\theta>1/2$ and $\theta<1/2$ the
behaviour of the two channels are simply interchanged.

\section{Steady state density profiles}
\label{steady}
\begin{figure}[h]
\centering
\includegraphics[width=7.5cm,bb=50 50 410 302]{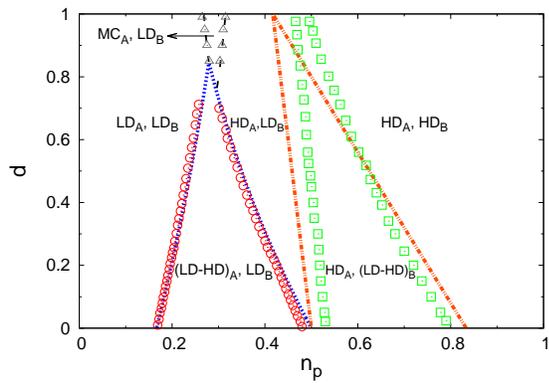}
\caption{\rr{(color online)} Phase diagram with $\theta=0.8$ for densities $\rho_A$ and
$\rho_B$ for channels $T_A$ and $T_B$ respectively obtained from our
MFT \rr{(dashed and dotted lines)} and MCS (circles, squares and triangles) analysis. Staying up to $d=1$, $T_A$ has
four phases (all marked by suffix A in the phase diagram),
LD,LD-HD,HD and MC phase, where as, $T_B$ has only three phases
(marked by suffix B) without any MC phase. The \rr{dotted} blue and the
\rr{dashed black} lines correspond to the phase boundaries obtained for $T_A$, whereas the \rr{dashed-dotted} orange line for $T_B$ from MFT calculations. The red
circles and grey triangles correspond to the phase boundaries
obtained from MCS for $T_A$ and the green squares for that of
$T_B$.} \label{fig:theta.8_phase}
\end{figure}

\begin{figure}[h]
\centering
\includegraphics[width=7.5cm]{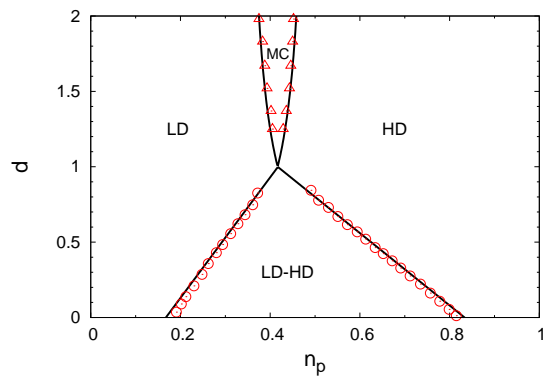}
\caption{\rr{(color online)} Phase diagram for both $T_A$ and $T_B$ for $\theta=1/2$.
There are four distinct phases; $T_A$ and $T_B$, being statistically
identical, have identical phase behaviour (see text). The LD-HD
phases in both $T_A$ and $T_B$ now have {\em delocalised} DWs.
Numerical data are shown by circles and triangles and the phase boundaries
obtained analytically are shown by solid lines.}
\label{fig:theta.5_phase}
\end{figure}

We use mean-field theory
(MFT) together with extensive Monte-Carlo simulation (MCS) of our
model to obtain the steady state density profiles. In the MFT, the
system is considered as a collection of three channels (two TASEP
and one SEP) with effective entry and exit rates~\cite{effective}. Once these
effective rates are determined from the condition of constancy of
particle currents, one may use them in conjunction with the known
results for TASEP and SEP with open boundaries to obtain the density
profiles here. Since an isolated TASEP in steady state can be in
three different states, the low density (LD), high density (HD) and
maximal current (MC) phases, and we have two active (TASEP)
channels, there are a number of possibilities for the overall
density profile
of the two active channels.
In order to ensure that the diffusive current does not vanish in the
thermodynamic limit (TL, see Ref.~\cite{frey}, see below also) we
let diffusivity $D$ scales with system size $L$ and define a
parameter $d=D/L$ which is the same for any arbitrary system size.
Thus steady states of the model are to be parametrised by
$(d,n_p,\theta)$.  Let us now set the notations:
\rd {for discrete lattice, density at a particular site $m$ is defined as
$\rho_i^m=\langle n_i^m \rangle$, where $i=A$ and $B$ refer to $T_A$
and $T_B$, and $i=S$ for the mean density in the SEP channel $S$. Further
in MFT considering continuum limit the density is defined
as $\rho_i(x)$, where $x=m/L,$ and in TL, $L\gg 1,$ $x$ lies in the
range $0\leq x\leq1.$ In all our MFT analysis we use the continuum labelling $x$ for the lattice in one dimension.}
Our main results are summarized in the phase diagrams
(parametrised by $n_p$ and $d$) for $\rho_A(x)$ and $\rho_B(x)$, as
shown in Fig.~(\ref{fig:theta.8_phase}), for a representative value
of $\theta=0.8$,  which are obtained by extensive Monte-Carlo
simulations of our $1d$ lattice-gas model and the corresponding $1d$
MFT; see below for details. \rd{The Monte-Carlo simulations
were realized by random sequential update}. For any value of
$\theta$ (except $\theta=0,1$, when one of the active channels is
closed), the phase diagrams of both $T_A$ and $T_B$ may display
a combination of the usual LD, HD and MC phases and a region of
co-existence of LD and HD phases, i.e., when the density profiles
show {\em localised domain walls (DW)}. The phase co-existence
regions are {\em non-overlapping} for $T_A$ and $T_B$, i.e., they do
not appear for the same values of $d$ and $n_p$ for a given
$\theta\neq 1/2$.
At $\theta=1/2$, for which $T_A$ and $T_B$ are statistically
symmetric, the density profiles $\rho_A(x)$ and $\rho_B(x)$ of
$T_A$ and $T_B$, respectively, are naturally identical. The
crucial difference with $\theta\neq 1/2$ is that the co-existence
region now corresponds to {\em delocalised DWs}, i.e., the fraction
of the system size $L$ visited by a fluctuating DW {\em does not}
vanish in TL $L\rightarrow\infty$. Such delocalised DWs appear in
{\em both} $T_A$ and $T_B$ only for $\theta=1/2$. Thus as
$\theta\rightarrow 1/2$, the model undergoes a {\em delocalisation
transition}. The phase diagram for $\theta=1/2$ are given in
Fig.~(\ref{fig:theta.5_phase}).

To begin with, we denote densities at the junction site by
\rd{$\alpha_{_{A,B}}=\rho_{A,B}(0)$ and $1-\beta_{_{A,B}}=\rho_{A,B}(1)$}
for the active channels, and
\rd{$\gamma=\rho_S(0)$ and $\delta = \rho_S(1)$}
in the passive SEP channel. The
SEP current then takes the well-known linear form,
\begin{equation}
 J_s=(\gamma-\delta)d.
 \label{eqn.SEP_current_1}
\end{equation}
Evidently, $J_s$ remains finite in TL provided $D$ scales with $L$
linearly, else, for a fixed $D$ the SEP current $J_s$ vanishes for
$L\rightarrow\infty$. This provides {\em a posteriori} justification
for the scale-dependent $D$ that we have mentioned before. Noting
that the current in each of $T_A$ and $T_B$ is given by $J_i =
\rho_i (1-\rho_i),\,i=A,B$ (assuming no boundary layer at
$i=A,B$, i.e., $T_A,\,T_B$ are in their LD/coexistence phases) and
using conservation of total current at the left and right
junctions for the individual incoming/outgoing currents
to/from $T_A$ and $T_B$ from/to the SEP
channel we obtain
\begin{equation}
J^{in}_A(0)=\delta(1-\aa)\theta D,\;J^{in}_B(0)=
\delta(1-\ab)(1-\theta) D.
\label{eqn.current_in_1}
\end{equation}
Next, the individual outgoing currents at the sites $1$ in $T_A$ and
$T_B$ are (again assuming that the channels are in LD or
coexistence phase)
\begin{equation}
J^{out}_A(0)=\aa(1-\aa),\;J^{out}_B(0)=\ab(1-\ab).
\label{eqn.current_in_2}
\end{equation}
Conservation of current then yields
$J^{in}_{_{A,B}}(1)=J^{out}_{_{A,B}}(1).$ As expected,
this holds so long as $T_A,\,T_B$ are in their LD or coexistence
phase.

In contrast, if $T_A,\,T_B$ are in their HD or coexistence
phases the total outgoing current from $T_A$ and $T_B$ to the SEP
channel is given by
\begin{equation}
J^{out}_{_{T}} (1) = (1-\ba)(1-\gamma) + (1-\bb)(1-\gamma).
\label{eqn.SEP_current_2}
\end{equation}
Conservation of total current at the right junction then yields
(assuming $T_A,T_B$ to be in HD or coexistence phases)
\begin{equation}
J^{out}_{_{T}} (1) = \ba(1-\ba) + \bb(1-\bb). \label{eqn.SEP_current_3}
\end{equation}

Further, again assuming HD or coexistence phases for
$\rho_{A,B}$, and separately considering the currents from
\rd{$T_{A,B}(x=1)$ to $S(x=0)$} yields $\ba=\bb$~\cite{hole}. This is
corroborated by our MCS simulations (see below). This
immediately yields that the bulk currents are equal. This is possible only
when the bulk
currents in $T_{A,B}$ are controlled by RJ, i.e., $T_A$ or $T_B$ are
both in HD or a combination of coexistence and HD phases. Notice
that the conditions obtained for \rd{$\rho_{A,B}(0)$ and
$\rho_{A,B}(1)$} by using current conservations at the respective
sites do not hold simultaneously, unless $T_{A}$ or $T_B$ are in
coexistence phases, such that there are no boundary layers at
\rd{$x=0,1$} with $\rho_{A,B}$ being piecewise continuous.
Having defined {\em effective} entry and exit rates (valid
separately for LD/coexistence or HD/coexistence phases in
$T_{A,B}$) for the active channels, we can now apply the known
results of TASEP here. One obtains the low (high) density phases in
the periodic system equally and are characterized by a uniform
density below (above) 1/2 and a boundary layer at the right (left).
However for $\alpha_{_{A,B}}=\beta_{_{A,B}}$, the boundaries are
matched by a piecewise constant density profile with an intervening
DW. For TASEP with open boundaries, particle entry and exit events
are uncorrelated, and as a result, the DW is delocalised and
undergoes random walks covering the entire span of the system in the
long time limit. However in the present model, as in
Ref.~\cite{frey}, entry and exit of particles are {\em not
uncorrelated}; they get correlated by the fact that the ends of the
active channels are connected by the passive channel. Consequently,
as revealed by our Monte Carlo simulation studies, we find localised
DW in the active channels, which is similar to Ref.~\cite{frey}.
However, rather surprisingly for the special case of $\theta=1/2$,
i.e., when each of the active channels carry equal current on
average, we obtain delocalised DWs in {\em both} channels. Our MFT
formulated above may now be used to analyse the density profiles in
the different channels of the model quantitatively.

\subsection{DW in one active channel and LD in other}
\begin{figure}[h]
\centering
\includegraphics[width=7.5cm,bb=50 50 410 302]{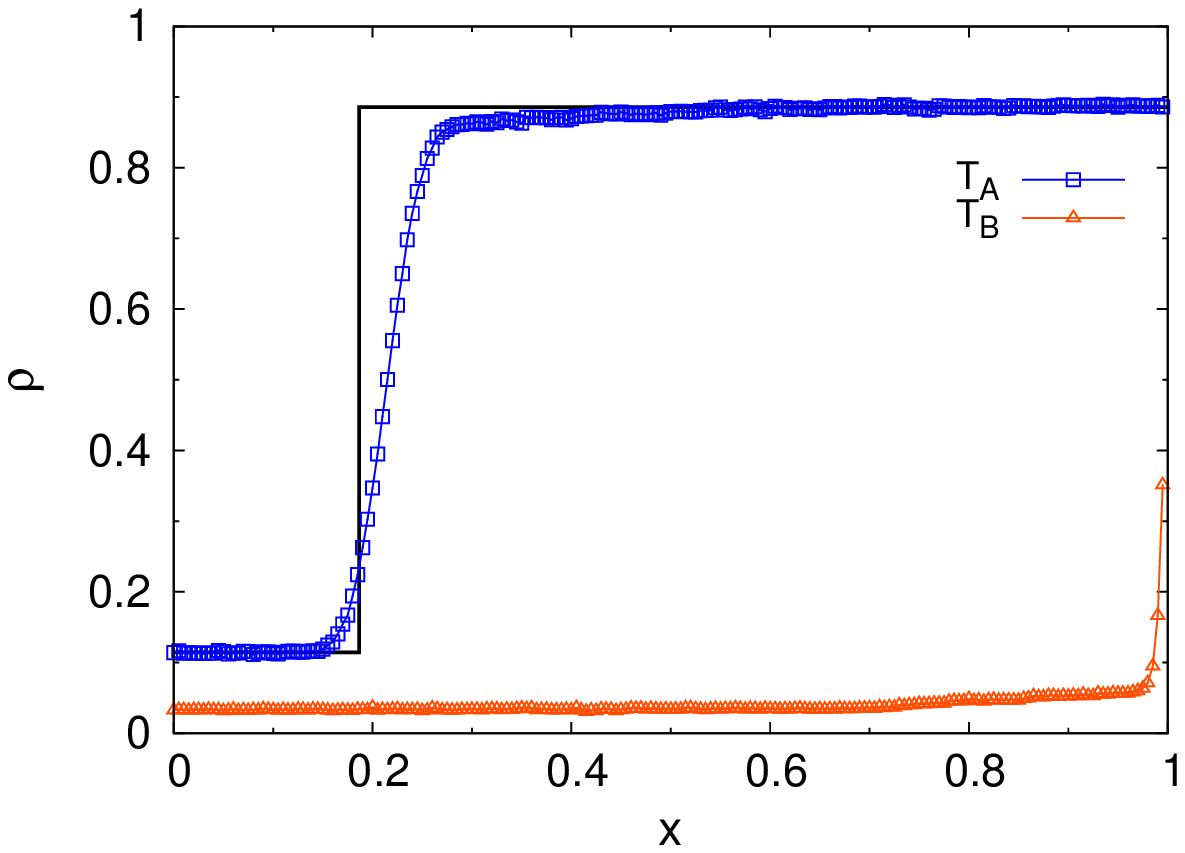}
\includegraphics[width=7.5cm,bb=50 50 410 302]{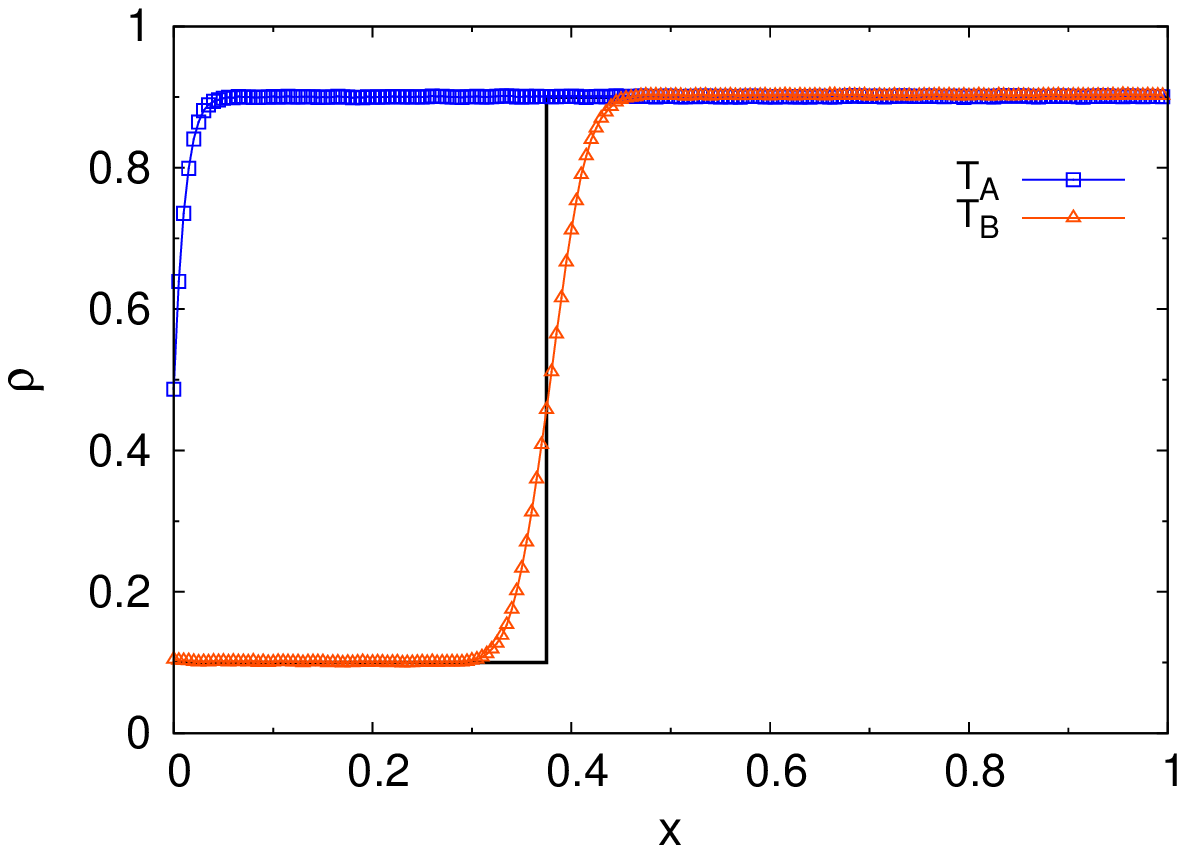}
\caption{ \rr{(color online)} (top) For $L=200$, $\theta=0.8$, $n_p=0.40$ and $d=0.15,$ 
$T_B$ is in LD phase and DW appears in $T_A$ at $x_w^A \sim 0.21,$
whereas from MFT Eq.(\ref{eqn.dwall}) $x_w^A=0.19.$ (bottom) For
$L=200$, $\theta=0.8$, $n_p=0.65$ and $d=0.20,$ $T_A$ is in HD phase
and DW appears in $T_B$ at $x_w^B \sim 0.38,$ \rd{ whereas from MFT
Eq.(\ref{eqn.dwall_B})  $x_w^B=0.375.$ The black continuous lines
display the domain walls obtained from MFT.}} \label{fig:dw_1}
\end{figure}

First, let us consider a situation when there is a DW in one of the active
channels (say $T_A$ with $\theta>1/2$) and the other active channel
($T_B$) is in the LD phase with a uniform density $\ab$ (within MFT
neglecting any boundary layer). Following the phenomenology of TASEP
with open boundaries, we set $\aa=\ba$ as a requirement of a DW in
$T_A$. Possibilities of simultaneous DWs in $T_A$ and $T_B$ will be
discussed later.  Within MFT, $\rho_A$ may be represented by a
Heaviside function that connects the two regions of constant density
through a localized DW at position $x_w^A$ (say) as,
\begin{equation}
\rho_A(x) = \aa + \Theta(x-x_{w}^A)(1-\aa-\ba).
\label{eqn.thetafunc}
\end{equation}
Since $T_B$ is assumed to be in the LD region, density $\rho_B(x)$
can be written as,
\begin{equation}
\rho_B(x) = \ab, \label{eqn.rhob}
\end{equation}
neglecting the boundary layer at the right boundary. For
the SEP channel the linear density distribution gives,
\begin{equation}
\rho(x) = \delta + (\gamma - \delta)x.
\end{equation}
Further, the particle number conservation can be expressed as
\begin{equation}
\label{eq-np} 3n_p = \int^{1}_0 dx [~ \rho(x) + \rho_A(x) +
\rho_B(x) ~],
\end{equation}
following the conditions as above and disregarding the
discontinuities at the right boundaries. Again from
Eq.(\ref{eqn.current_in_1}) and Eq.(\ref{eqn.current_in_2}) we have,
\begin{equation}
\ab = \aa\left(\frac{1}{\theta} - 1\right) =\aa q,
\label{eqn.2alpha}
\end{equation}
where $q=(1/\theta - 1).$ Now by solving Eq.~(\ref{eq-np}) in TL
\rd{ ($\delta \rightarrow 0$)} we get
\begin{equation}
x_w^A = \frac{1+ \frac{\gamma}{2}- 3n_p -\aa(1-q)}{1-2\aa}.
\label{eqn.dwall}
\end{equation}
Again  Eqs.~(\ref{eqn.SEP_current_2}), (\ref{eqn.SEP_current_3}),
(\ref{eqn.2alpha}) \rd{ and the relations $\ba=\aa$ and $\bb=1-\ab$}
yield for $\gamma$ as,
\begin{equation}
 \gamma=\frac{\aa^2(1+q^2)-2\aa+1}{1-\aa(1-q)}.
 \label{gamma_expression}
\end{equation}
In TL $\delta \rightarrow 0$ and $\gamma = J_{_{s}}/d.$ Since the
model considered here is closed,
$J_s= J_A^{out}(0) + J_B^{out}(0)$. Again from
Eq.~(\ref{eqn.current_in_2}) we have,
\begin{equation}
d=\frac{\aa(1-\aa)+(1-q\aa)q\aa}{\gamma} \label{eqn.d_gamma_1}.
\end{equation}
Hence, the position of the DW depends on the two control parameters
$n_p$ and $d$ for a given $\theta$. When the DW in $T_A$ is
localised within the system ($0<x_w^A<1$), it connects the LD and HD
phases of $T_A$ through a phase of coexistence (LD-HD). The
boundaries between the LD, LD-HD phases and LD-HD, HD phases of
$T_A$ are obtained by setting $x_w^A=0$ and $x_w^A=1$ respectively.
Setting $x_w^A=0$ from Eq.(\ref{eqn.dwall}) we get a quadratic
equation in $\aa.$ For a particular value of $n_p$ the feasible
values will be $(0<\aa<1/2).$ Now putting that $\aa$ in
Eq.~(\ref{eqn.d_gamma_1}) we get the corresponding $d$ value. Thus
we get the boundary between LD and LD-HD coexistence phase in the
$(n_p,d)$-plane. Similar exercise for $x_w^A=1$ gives the right
boundary between LD-HD and HD phase. See
Fig~(\ref{fig:theta.8_phase}) for details.  \rd{ In
Fig~(\ref{fig:theta.8_phase}) the phases of channel $T_A$ and $T_B$
are spanned by $d\le1$ and $n_p\le1$. The phase diagrams for $T_A$
and $T_B$ are drawn corresponding to a situation when $T_A$ displays
a variety of phases (LD,LD-HD,HD AND MC), while $T_B$ remains in its LD
phase. For $d>1$, this part of the phase diagram remains
qualitatively unchanged, with the phase marked as $MC_A,LD_B$ (i.e.,
$T_A$ in MC and $T_B$ in LD) should expand to a larger area.
Similarly, the phases of $T_B$ are shown when channel $T_A$ remains
in its HD phase. The condition for the latter is mathematically
given by $\rho_A(x)=1-d/2$ in the bulk. For $1<d<2, \rho_A(x) <1/2$
and hence $T_A$ is no longer in its HD phase. Hence, we do not
consider the $d>1$ region while presenting our phase diagram.} A DW
in $T_A$ obtained from our MCS studies are shown in
Fig.~(\ref{fig:dw_1}, top) with $\theta=0.80$, $n_p=0.40$ and
$d=0.15$. \rd{ We have taken $L = 100$ and $200$ for determining the
position of the domain walls and phase diagram. We do not find any
significant dependence of our results on $L$.}

\subsection{DW in one active channel and HD in other}
Let us now consider the case when there is a DW in $T_B$, and $T_A$ is in
HD phase for $\theta > 1/2$, thus having a boundary wall
at the left end. As discussed above, within MFT, $\rho_B$ may be
represented by Heaviside $\theta$-function as
\begin{equation}
\rho_B(x) = \ab + \Theta(x-x_{w}^B)(1-\ab-\bb),
\label{eqn.thetafuncB}
\end{equation}
and $T_A$ is in HD phase having a uniform density of $(1-\ba)$,
neglecting the boundary layer. To have a DW in  $T_B$ we
must have $\ab=\bb$. Now from Eqs.~(\ref{eqn.SEP_current_2}) and
(\ref{eqn.SEP_current_3}), we get
\begin{equation}
\gamma = \frac{(1-\ba)^2 + (1-\bb)^2}{2-\ba-\bb}.
\label{gamma_expression0}
\end{equation}
Hence Eq.~(\ref{gamma_expression0}) and $\ba = \bb$ yield $\gamma =
(1 - \ab)$ . In TL $\gamma = J_s/d$, and hence, $\ab = d/2$. Again
by using the particle number conservation and \rd{ as $\ba=\bb=\ab$ and
$\rho_A(x) = 1 - d/2$} we obtain,
\begin{equation}
x_w^B = 1- \frac{3n_p - \frac{3}{2} + \frac{d}{4}}{1-d}.
\label{eqn.dwall_B}
\end{equation}
From the above expression we get the boundaries between the LD,
LD-HD ($x_{w}^B=0$) and LD-HD, HD phases ($x_{w}^B=1$) of $T_B$. A
DW in $T_B$, obtained in our MCS studies, is given in
Fig.~(\ref{fig:dw_1}, bottom). There is a crucial difference between
the DWs in $T_A$ and $T_B$: The LD part of the DW in $T_A$ has
density $\aa$, different from the density $\ab$ of $T_B$ (fully
in LD), thus $\rho_A(x)$ has no overlap with $\rho_B(x)$. In
contrast, $\rho_A(x)$ (fully in HD) overlaps with $\rho_B(x)$
in the HD part of the DW in $T_B$. This is due to $\ba=\bb$ and is
clearly visible in Fig.~(\ref{fig:dw_1}).

\subsection{Delocalised domain wall at $\theta=1/2$}
Let us now carefully consider the properties for $\theta=1/2$, when
both the active channels are symmetric and statistically identical.
Thus, if $\aa=\ba$ then automatically $\ab=\bb$. Hence, if $T_A$ has
a DW, $T_B$ too will have a DW, or is in its LD-HD (co-existence)
phase as well, such that its density may be represented by a
Heaviside $\theta$-function that connects the two regions of
constant density through a localized DW at $x_w^B$ (say). Hence,
$\rho_A(x)$ and $\rho_B(x)$ are given by the expressions
(\ref{eqn.thetafunc}) and (\ref{eqn.thetafuncB}) respectively. As
both  $T_A$ and $T_B$ show DWs, thus $\aa=\ba$ and $\ab=\bb,$ and
for $\theta=1/2$ from Eq.(\ref{eqn.current_in_1}) we have,
$\aa=\ba=\ab=\bb=\alpha$ (say). Again, from
Eq.(\ref{eqn.current_in_2}) and Eq.(\ref{eqn.SEP_current_2}) we have
$\gamma = 1-\alpha.$ Therefore, in TL, $\delta \rightarrow 0$ and
thus $\alpha=d/2.$ Now by particle number conservation we have,
\begin{equation}
3n_p = 1/2 + 3d/4 + (2-x_{w}^A-x_{w}^B)(1-d).
\label{Eq.particle_conserve_2}
\end{equation}
Thus for $\theta=1/2$ we get a relation given by
Eq.(\ref{Eq.particle_conserve_2}) between $x_{w}^A$ and $x_{w}^B$
for a particular value of $n_p$ and $d.$ In other words $x_w^A$ and
$x_w^B$ are {\em not} uniquely determined. In this case both DWs are
delocalized and perform random walk along the active channels. As
$T_A$ and $T_B$ are identical,  the condition for DW is satisfied
for both the channels simultaneously. Thus, $x_w^A=x_w^B=0$ and
$x_w^A=x_w^B=1$ in Eq.(\ref{Eq.particle_conserve_2}) can give the
boundaries of LD-HD phase in both channels with the LD and HD phases
respectively. In Fig.~(\ref{fig:theta.5_phase}) we have shown this
mean-field result as well as that obtained from MCS which shows
distinct four phases with phase boundaries for both channels are
identical. However, symmetry between the two active channels dictate
that the long-time
\begin{figure}[h]
\centering
\includegraphics[width=7.5cm,bb=50 50 410 302]{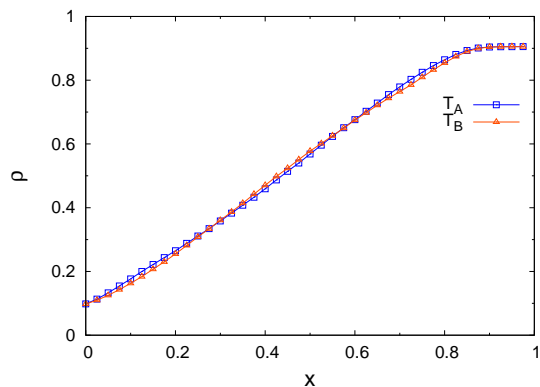}
\caption{ \rr{(color online)} Plots of $\rho=\rho_A,\,\rho_B$ versus site (x) for $\theta=1/2$
with $n_p=0.52$ , $d=0.19$ and $L=2\times10^3$. 
Clearly, both $\rho_A$ and $\rho_B$  display overlapping delocalised
DWs.} \label{fig:dw_both}
\end{figure}
averaged positions of the DWs (equivalently, the long-time average
density profiles $\rho_A(x)$ and $\rho_B(x)$) in $T_A$ and $T_B$ are
identical, a fact verified by our MCS, is displayed in
Fig.~(\ref{fig:dw_both}).

\subsection{Both the active channels in MC phase}
 Lastly, we consider the possibility of the
MC phases in the active channels. Let us first consider the
conditions for obtaining MC phases in both the active channels.
\rd{ Condition for MC in isolated TASEPs are $\rho_A = \rho_B = 1/2$ and
$J_A=J_B=1/4$, and this happens when all the boundary densities
$\alpha_A, \alpha_B, \beta_A, \beta_B > 1/2$.}
Furthermore, $T_A,\,T_B$
have boundary layers at both the ends. This precludes usage of
 Eq.~(\ref{eqn.current_in_2}) to determine the boundary densities. Using
$J_A=1/4=J_B$ together with Eq.~(\ref{eqn.current_in_1}) (assuming
no density discontinuities between \rd{ $\rho_S(1)$ and $\rho_{A,B}(0)$}), we find $\delta (1-\alpha_A)\theta D=1/4,\,\delta (1-\alpha_B)
(1-\theta)D=1/4$. Since $\alpha_{A,B}>1/2$
\begin{equation}
\delta > max\left\{\frac{1}{2dL\theta},\frac{1}{2dL(1-\theta)}\right\}
\label{delta_greater_MC2}
\end{equation}
Using similar considerations at RJ, and again assuming no
density discontinuity between \rd {$\rho_S(0)$ and $\rho_{A,B}(1)$}, which means
$1-\gamma = \beta_A = \beta_B$, together
with  $\ba,\bb
> 1/2$,  Eq.~\ref{eqn.SEP_current_1} and $J_s = 1/2$ we have,
\begin{equation}
 \delta < \frac{1}{2} - \frac{1}{2d}.
 \label{delta_lesser_MC2}
\end{equation}
From particle conservation we have,
\begin{equation}
 3n_p = \delta + \frac{1}{4d} + 1
 \label{eqn.inequality_MC2}
\end{equation}
Eqs.~(\ref{delta_greater_MC2}), (\ref{delta_lesser_MC2}) and
(\ref{eqn.inequality_MC2}) yield boundaries of the MC phase with LD
and HD phase as, $d=1/(12n_p-4)$ and $d=1/(6-12n_p)$ which indicates
the presence of such phase for $d>1$ and bounded by the two above
mentioned lines. Thus, the demarcating lines are independent of
$\theta$. They are shown in Fig.(\ref{fig:theta.5_phase}). We now
consider the case when one of the channels (say, $T_A$) is in MC
phase and the other one ($T_B$) in the LD phase. Therefore, we have
$J_{A}=1/4$, $\rho_A=1/2$ and $\aa,\ba>1/2.$ From
Eqn.~(\ref{eqn.current_in_1}) and (\ref{eqn.current_in_2}) we have
$\aa=D\delta \theta.$ Again using the MC phase condition
$\aa,\beta_A>1/2$ we have (arguing as before),
\begin{equation}
 \delta > \frac{1}{2dL\theta},\,\delta < \frac{1}{2} - \frac{J_s}{d}.
 \label{delta_greater}
\end{equation}
The maximal current condition gives $J_s=(1/4 + q/2 -q^2/4),$ then from particle conservation we have,
\begin{equation}
 3n_p = \delta + \frac{J_s}{2d} + \frac{1}{2} + \frac{q}{2}
 \label{eqn.inequality}
\end{equation}
The two inequalities (\ref{delta_greater}) together with
Eq.(\ref{eqn.inequality}) then yield boundaries of the MC phase with
the LD and HD phase respectively as, $d=J_s/(6n_p-q-1)$ and
$d=J_s/(q+2-6n_p)$. In Fig.~(\ref{fig:theta.8_phase}) we have shown
the MC phase boundaries. Our MCS studies also reveal a small MC
phase within the region obtained from MFT. Not surprisingly, for
$\theta=1$ and $\theta=0$, the MC phase regions obtained from our
MFT match exactly with that of Ref.~\cite{frey}. In addition, one
may argue that the coexistence of $T_A$ in HD  and $T_B$ in MC  is
not possible. For $T_A$ to be in HD phase $1-\ba > 1/2$ or $\ba <
1/2$. Again  $\bb = \ba < 1/2$, so long as $T_{A,B}$ are
in HD or coexistence phases. But condition for MC phase in $T_B$ is
$\ab,\bb
> 1/2$. Thus, an MC phase in $T_B$ (when $T_A$ in HD) is not allowed.

\section{Domain wall fluctuations and delocalisation transition}
\label{local} Until now we have considered the MFT for the model,
where all fluctuations are neglected. However, the DWs fluctuate
about their MF DW (mean) positions $x_w^A$ or $x_w^B$. We have
studied these fluctuations numerically and characterise them by
measuring the scaling of the fluctuations with $\theta$ and system
size $L.$
\begin{figure}[h]
\centering
\includegraphics[width=7.3cm]{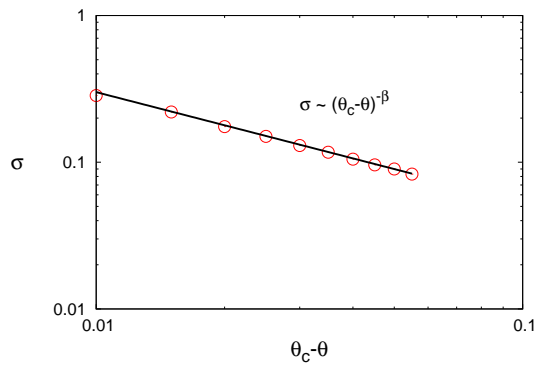}
\caption{\rr{(color online)} Log-log plot of DW width $\sigma$ versus
$|\theta-\theta_c|$ for $L=200$, $n_p=0.40$ and $d=0.15$
with the exponent $\beta=3/4$ (see text).}
\label{fig:delocalization_transition}
\end{figure}
\begin{figure}[h]
\centering
\includegraphics[width=7.3cm]{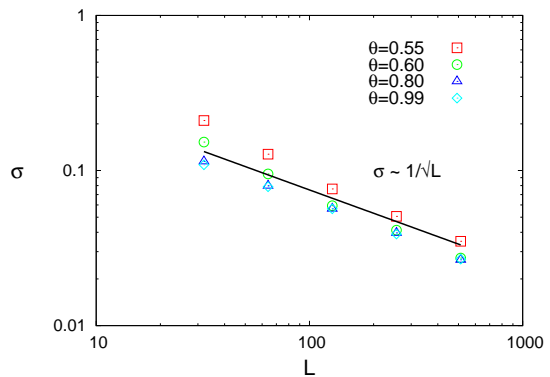}
\caption{\rr{(color online)} Log-log plot of $\sigma$ versus  $L$ for $n_p=0.40$ and $d=0.15$ at an {\em off-critical point:} $\theta\neq \theta_c$. The line has a slope
1/2; thus, $\sigma\sim 1/\sqrt L$ (see text).} 
\label{fig:dw_width}
\end{figure}
In particular as $\theta\rightarrow \theta_c=1/2$ from
above or below, the localised DW in $T_A$ or $T_B$ shows a {\em
delocalisation transition} at which DW fluctuations diverge.
The width $\sigma$ of the distribution of DW fluctuations can be
obtained by fitting the density profile in the vicinity of the
domain wall by the function $(P\cdot erf[(x-Q)/\sigma]+R)$
\cite{reich}, with the parameters $P, Q, R, \sigma.$ We find
$\sigma$ to diverge with  a power law dependence on $(\theta -
\theta_c)$ as,
\begin{equation}
\sigma \sim (\theta_c-\theta)^{-\beta} \label{eqn.theta_c},
\end{equation}
with $\beta=3/4$ obtained from our MCS studies as shown in
Fig.~(\ref{fig:delocalization_transition}).
In contrast, at an {\em off-critical point}, i.e., for $\theta\neq 1/2$, DW fluctuations are finite and vanish in TL $L\rightarrow\infty$ as $L^{-1/2}$; 
we have shown this in Fig.~(\ref{fig:dw_width}). \rd{ For investigating the variation of domain wall width with $L$ for various values of $\theta$ we have taken $L = 32, 64, 128, 256$ and $512$.}

\section{Summary and outlook}
\label{conclu} Analytical and numerical studies of our model amply
illustrate the underlying rich phase behaviour, including a
delocalisation transition, unexpected in a system without
boundaries. While boundary-induced phase transitions including
delocalisation transitions have been observed in several open
systems with exclusion processes together with spatially nontrivial
steady state densities~\cite{reich,lktasep,gunter}, analogous
studies on bulk closed systems are less studied so far. The
competition between the diffusive and driven dynamics, and the
division of the SEP current into two parallel TASEP currents are
crucial to the macroscopic behaviour we obtained. The latter is
controlled by a parameter $\theta$, which is a tuning parameter in
the model. \rd{ The most striking feature in our work vis-a-vis the
results in Ref.~\cite{frey} is the possible existence of a
delocalisation transition and correspondingly the formation of DWs
in both $T_A$ and $T_B$ simultaneously at a special value
$\theta=1/2$. In contrast to the DWs formed either in $T_A$ or $T_B$
(but not simultaneously in both) for $\theta\neq 1/2$, as found in
Ref.~\cite{frey} as well as in the present work, the DWs at
$\theta=1/2$ are no longer pinned to a fixed point in the lattice
with vanishing fluctuations in the thermodynamic limit. Instead they
delocalise and have position fluctuations that do not vanish in the
thermodynamic limit. Thus the parameter $\theta$ in our model
appears as a tuning parameter or a switch, which can be used to
control the nature of domain wall fluctuations
(localised/delocalised). In addition for $\theta \neq 1/2,$ the value of
$\theta$ can be tuned to make the DW appear or disappear in one of the 
active channels. There is no analogue of these in the study of Ref.~\cite{frey}.} 
While we have considered only two TASEP
channels, many more may be added and studied systematically as
above. Recalling protein synthesis by ribosomes along mRNA strands
as one of the phenomenological motivation for our model, it may be
noted that several mRNAs compete for same resources (ribosomes) in a
cell. Thus a systematic study of multiple TASEP channels connected
in parallel with a single SEP channel would be useful. The failure
of the traditional MFT calls for further analysis by means of more
sophisticated analytical techniques, e.g., Bethe ansatz
\cite{gunter} or density matrix renormalisation group \cite{dmrg},
which are beyond the scope of the present work. From the point of
view of nonequilibrium statistical mechanics, our model belongs to
the class of models lacking translation invariance and without
boundaries that displays a phase transition (in the form of a
delocalisation transition). Our model may be extended in several new
directions, e.g., again motivating by ribosome movements along mRNA,
one may in our model consider particle exchanges between $T_A$ and
$T_B$, or between one of the active channels and passive channel
(representing ribosome attachments or detachments),  allow defects
along the active channels (representing defects in the mRNA),
introduce a second control parameter at the exit ends of $T_A$ and
$T_B$ that controls the relative outgoing currents to SEP and
unequal hopping rates in $T_A$ and $T_B$. These will be considered
elsewhere. \rr{ We close this work with a note of caution: As mentioned in the beginning, despite some similarities our model cannot be directly used for quantitative descriptions of ribosome translocations along mRNA strands due to its limiations. First of all, ribosome diffusion takes place inside a cell, which, alothough geometrically confined, has a three-dimensional ($3d$) structure, as opposed to our $1d$ diffusive model for it. Secondly, the description of a ribosome
as a {\em single unit} (i.e., a point particle here) is also questionable, for it gets released from an mRNA by falling apart into different subunits, a feature not possible to capture in our simplified description here. Nevertheless, our work provides some clues about the actual biological system, e.g., the crucial role of particle number conservation in determining the nature of the steady states. We expect that more realistic theoretical descriptions of ribosome translocation and detachment should have some of the basic features of our model in-built into it.}
\acknowledgments AB wishes to thank the Max-Planck-Gesellschaft
(Germany) and Department of Science and Technology (India) for
partial financial support through the Partner Group programme
(2009). AKC acknowledges the financial support from DST (India)
under the SERC Fast Track Scheme for Young Scientists [Sanction no.
SR/FTP/PS-090/2010(G)].

\end{document}